\documentclass[fleqn,10pt]{wlscirep}
\usepackage[utf8]{inputenc}
\usepackage[T1]{fontenc}
\usepackage{mathtools}
\DeclarePairedDelimiter{\ceil}{\lceil}{\rceil}

\title{Stochastic diffusion using mean-field limits to approximate master equations}

\author[1,2,*]{Laurent H\'ebert-Dufresne}
\author[3,4]{Matthew M. Kling}
\author[1,2]{Samuel F. Rosenblatt}
\author[5,6]{Stephanie N. Miller}
\author[3]{P. Alexander Burnham}
\author[1,7]{Nicholas W. Landry}
\author[3,4]{Nicholas J. Gotelli}
\author[5,6]{Brian J. McGill}
\affil[1]{Vermont Complex Systems Center, University of Vermont, Burlington VT, USA}
\affil[2]{Department of Computer Science, University of Vermont, Burlington VT, USA}
\affil[3]{Department of Biology, University of Vermont, Burlington VT, USA}
\affil[4]{Gund Institute for Environment, University of Vermont, Burlington VT, USA}
\affil[5]{Mitchell Center for Sustainability Solutions, University of Maine}
\affil[6]{School of Biology and Ecology, University of Maine}
\affil[7]{Department of Biology, University of Virginia, Charlottesville, VA, USA}

\affil[*]{laurent.hebert-dufresne@uvm.edu}

\begin{abstract}
Stochastic diffusion is the noisy and uncertain process through which dynamics like epidemics, or agents like animal species, disperse over a larger area. Understanding these processes is becoming increasingly important as we attempt to better prepare for potential pandemics and as species ranges shift in response to climate change. Unfortunately, modeling of stochastic diffusion is mostly done through inaccurate deterministic tools that fail to capture the random nature of dispersal or else through expensive computational simulations. In particular, standard tools fail to fully capture the heterogeneity of the area over which this diffusion occurs. Rural areas with low population density require different epidemic models than urban areas; likewise, the edges of a species range require us to explicitly track low integer numbers of individuals rather than vague averages. In this work, we introduce a series of new tools called ``mean-FLAME" models that track stochastic dispersion using approximate master equations that explicitly follow the probability distribution of an area of interest over all of its possible states, up to states that are active enough to be approximated using a mean-field model. In one limit, this approach is locally exact if we explicitly track enough states, and in the other limit collapses back to traditional deterministic models if we track no state explicitly. Applying this approach, we show how deterministic tools fail to capture the uncertainty around the speed of nonlinear dynamical processes. This is especially true for marginal areas that are close to unsuitable for diffusion, like the edge of a species range or epidemics in small populations. Capturing the uncertainty in such areas is key to producing accurate forecasts and guiding potential interventions. Because it can model this uncertainty at a much lower computational cost than full master equations or exact simulations, we hope that our new modeling approach will enable more creative research to guide necessary adaptations for rural and marginal areas against the uncertain impacts of future pandemics and climate change.
\end{abstract}
\begin{document}

\flushbottom
\maketitle

\thispagestyle{empty}

\section{Introduction}
Self-organizing structures (e.g., populations, metapopulations, communities) and their complex dynamics (e.g., epidemics, species diffusion, culture) involve a combination of deterministic trends and stochastic effects. Although most ecological modeling efforts have focused on the deterministic component, the stochastic component can be critical when considering the movement of entities through time and space \cite{mendez2014biological}. Models of spread or diffusion, in particular, often rely on approximations that are built from deterministic assumptions \cite{Diekmann1995, peterson2002future}, like supposing well-mixed populations \cite{balcan2009multiscale}, or by simply tracking the average state of a system \cite{Kermack1927, pacala1995details}. These approximation methods work well if the system is large enough or if the probabilities of rare events are low enough. Still, they often do a poor job representing non-deterministic processes near the limits of systems (e.g., small populations, populations near the edge of their geographic range). Here, we present a novel method that can better approximate the state of ecological systems under near-limit conditions. 

First, consider the spread of an emergent disease under differently structured populations. In a large, well-connected population, a virulent disease could spread quickly, moving among dense population centers and eventually among sparser rural areas once transmission has exceeded a high enough rate \cite{cuadros2021dynamics}. Alternatively, in a network of small, isolated populations (e.g., bark beetles moving among stands of their host trees), transmission to new areas occurs more slowly, if at all, and relies on rare events. Models that ignore stochasticity will capture the dynamics of the large connected population but fail to capture the sparse stochastic transmission chains critical to the small isolated populations. With close to half of the global population living in rural areas, this is a large omission for many state-of-the-art epidemic models and the ecology of emerging diseases \cite{wilcox2005disease}.

Second, under climate change, organisms are expected to track suitable conditions by shifting their geographic ranges \cite{kerr2020racing}. For the range of a species to shift, individuals need suitable conditions to effectively survive and reproduce, and need to reach locations with those conditions. In cases where distances are short and there are no physical barriers between current occurrences and future suitable locations, modeling these range shifts using deterministic mean-field averages may be appropriate. However, in cases where long-distance dispersal is required to reach suitable conditions or get beyond barriers (e.g., mountain ranges, rivers, highways), stochastic events of dispersal are more important. Rare and random dispersal events have shaped species ranges as we know them today \cite{cain1998seed, clark1998trees}. Capturing rare and random dispersal is therefore critical to understanding how ecosystems will shift in response to climate change \cite{clark1998reid, travis2013dispersal}. To better represent in a modeling context the stochastic components like those critical to species range shifts and the spread of emergent diseases, we propose using a new approach based on the master equation framework.

Master equations are systems of equations that track all possible states and all possible transitions between states of a system. In theory, the application of master equations creates an exact description of the probability distribution over the possible states of a system through time. However, for spatially embedded systems, this approach is impractical because there are simply too many states to track given the large number of possible states for each location. Therefore, to make master equations applicable to spatial problems, we propose two transformations: Simplifying the descriptions of local states and approximating correlations to avoid tracking complex joint distributions. We do so by introducing \textit{mean-field limits to approximate master equations} (or mean-FLAME).

The rest of the paper is structured as follows. We first introduce the concept of master equations and mean-field limits without space, using a simple birth-death process for a single population and a coupled two-population system, in Sec.~\ref{sec:birthdeath}. We then extend this method to a two-dimensional process using Lotka-Volterra predator-prey dynamics, again without space, in Sec.~\ref{sec:LV}. We then apply the method to spatially embedded multidimensional examples in Secs.~\ref{sec:epi} and \ref{sec:dispersal}, using the above scenarios of metapopulation epidemic dynamics and species dispersal, respectively. We conclude by sharing some details of our software implementations of mean-FLAME and suggesting improvements for future work.

\section{Modeling approach with a birth-death process example \label{sec:birthdeath}}

In this section, we use a stochastic birth-death process for a single population to describe three components of our proposed methodology. First we present the traditional master equation framework which tracks the probability distribution of a system over all of its possible states by accounting for transition probability between discrete states. Next, we use a mean-field limit to approximate the master equation far from any absorbing state where dynamics die (for population size far above zero in this birth-death example). Then we show how approximating correlations across sub-systems or locations can simplify the processing of the master equation for coupled or spatially embedded systems. 

For this example we are interested in the number $n$ of individuals found in a population (system) at time $t$. In this example births and deaths are stochastic events that follow a Poisson process with births occurring at a rate $\mu n$ and deaths at rate $\nu n^2$. With this formulation, $n^{*}=0$ is an absorbing state that systems cannot leave (i.e., extinction) and there exists a meta-stable state at $n^{**} = \mu/\nu$. 

In a classic mean-field description, we would track the expected number of individuals $\bar{n}$ at time $t$. This would traditionally be done with the following ordinary differential equation:
\begin{equation}
    \frac{d}{dt}\bar{n}(t) = \mu \bar{n}(t) - \nu \bar{n}(t)^2 \; . \label{eq:mf}
\end{equation}
With this equation, the absorbing state $n^{*} = 0$ is always unstable if $\mu > 0$ such that any system with initial conditions $\bar{n}(0)>0$ will end up in the stable state $n^{**} = \mu / \nu$. However, consider an actual stochastic simulation of a system with $n=1$ active individual; there is a probability $\nu/(\mu+\nu)$ that the system falls into the absorbing state instead of moving to $n=2$. Mean-field approximation do not capture these stochastic extinction events. In the ecological literature, stochastic outcomes like this are referred to as demographic or environmental stochasticity and can be important drivers of population change \cite{engen1998demographic}. In fact, all populations could eventually fall into the absorbing state, and the rate at which these extinction events occur will depend on how close $n^{**} = \mu / \nu$ is to zero. We wish to capture these different paths to extinction.

\subsection*{Master equation}

Master equations are systems of equations that track the \textit{occupation probability} (or occupation number) for every possible state of a system by calculating the exact transition rates between states\cite{haag2017modelling}. Consider $P_{\overrightarrow{y}}(t)$, the occupation probability of state $\overrightarrow{y}$ at time $t$. The state $\overrightarrow{y}$ in this example is a simple count of individuals but it could include any description of the state of the system (e.g., location, epidemiological status or age of individuals, local policies), as long as $\overrightarrow{y}$ is sufficient to exactly specify the transition rates $\omega(\overrightarrow{y},\overrightarrow{z})$ from state $\overrightarrow{y}$ to other states $\overrightarrow{z}$ and vice versa. We can then define a general master equation for $P_{\overrightarrow{y}}(t)$ as, 

\begin{equation} \label{eq:general_me}
    \frac{d}{dt}P_{\overrightarrow{y}}(t) = \sum _{\overrightarrow{z}}\omega(\overrightarrow{z}, \overrightarrow{y}) P_{\overrightarrow{z}}(t) - \sum_{\overrightarrow{z}} \omega(\overrightarrow{y}, \overrightarrow{z}) P_{\overrightarrow{y}}(t),
\end{equation}
where positive terms correspond to probability flowing into state $\overrightarrow{y}$, from $\overrightarrow{z}$, and negative terms to probability flowing from $\overrightarrow{y}$ to other states. Often, the sums in Eq.~(\ref{eq:general_me}) involve only a few terms as the transition matrix across all states can be quite sparse. The complexity of the approach is instead related to the number of states available to the system, since the approach is only exact if all states are tracked explicitly.

In general, the master equation describing a given stochastic process can be written by answering a simple set of questions.
\begin{enumerate}
    \item What is the set of discrete states available to the system? The answer to this questions fixes the number the number of variables and differential equations needed.

    \item What are the possible transitions between states? The answer to this questions dictates how many terms with other states $\overrightarrow{z}$ should appear in a given differential equation.

    \item At what rates do these transitions occur?  These rates fix the transition factors $\omega$ in the master equation.
\end{enumerate}

To write the master equation of this birth death process, we follow our simple recipe. The system consists of a discrete number of active individuals $n$ which can take any value from 0 to infinity. The system can go from state $n$ to $n+1$ through reproduction or to state $n-1$ when an individual dies. The reproduction transition from $n$ to $n+1$ occurs at rate $\omega(n,n+1) = \mu n$ and the death transition to $n-1$ occurs at rate $\omega(n,n-1) = \nu n^2$ for all possible values of $n$; all other transitions are impossible. The corresponding master equation is therefore
\begin{equation}
    \frac{d}{dt}P_n(t) = (n+1)^2 \nu P_{n+1}(t) + (n-1) \mu P_{n-1}(t) - (n\mu + n^2\nu)P_n(t) \; \qquad \forall\; n\; \in\; \mathbb{N}. 
\end{equation}

\subsection*{Master equation with mean-field limit}

One possible issue with the master equation framework is that we need to track all possible states for the system. What if the population of active individuals goes to infinity? We can track thousands of equations without problem, but systems of several millions or billions of equations are computationally expensive.

\begin{figure}
    \centering
    \includegraphics{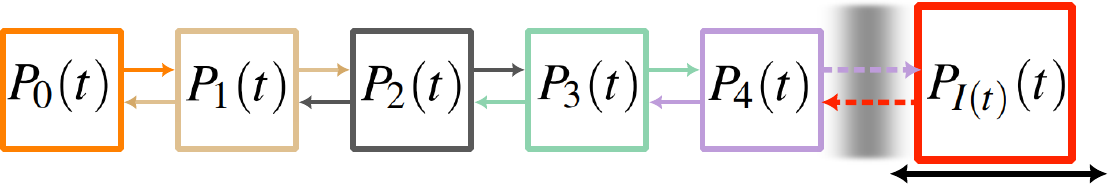}
    \caption{\textbf{Master equation with mean-field limit for a simple birth-death process.}}
    \label{fig:birth-death}
\end{figure}

We, therefore, introduce a mean-field limit in our master equation. We collapse all states with \textit{intense} activity as a mean-field quantity $I(t)$. Technically, we define intense activity as a system far enough from any absorbing state. In this birth-death example, this would be any population with size greater than some condition, $n\geq n_c$, far from extinction. For the approximation to be accurate, $n_c$ just needs to not be too close to zero. We then track both the occupation $P_{I(t)}(t)$ of that mean-field quantity and as well as its value $I(t)$. In our birth-death example, $I(t)$ represents the expected population size for populations with at least $n_c$ individuals, while $P_{I(t)}(t)$ represents the probability that a given population has at least $n_c$ individuals at time $t$. This approximation is illustrated in Fig.~\ref{fig:birth-death}. 

To include this mean-field limit in our master equation framework, we need to track the transitions in and out of that mean-field limit by assuming some distribution around its average state $I(t)$ for the fraction $P_{I(t)}(t)$ of systems therein. Indeed, based on whatever distribution we assume around the average state, we can calculate the probability that a system in the mean-field limit is actually right on the edge of the limit at $n=n_c$. Given that probability, we calculate the rate at which a system might leave the mean-field limit back to an explicitly tracked state at $n = n_c - 1$ in the master equation.

Let us write the system of equations for the birth-death process using the scheme shown in Fig.~\ref{fig:birth-death} where we use a mean-field $I(t)$ to describe systems with $n_c = 5$ or more active individuals (which is an arbitrary choice for the mean-field condition $n_c$). We need four types of equations: general master equation states, the last master equation state at $n_c - 1$ which is coupled to the mean-field limit, the occupation probability or occupation number of the mean-field limit, as well as the position of the mean-field limit itself. In the same order, these four types of equations can be written as follows for the birth-death process:
\begin{eqnarray}
    & \frac{d}{dt}P_n(t) &= (n+1)^2 \nu P_{n+1}(t) + (n-1)  \mu P_{n-1}(t) - (n \mu + n^2\nu)P_n(t) \qquad \forall\; n\; \in\; [0,n_c-2] \\
    & \frac{d}{dt}P_{n_c-1}(t) &= n_c^2 \nu \rho\left[n_c,I(t)\right] P_{I(t)}(t) + (n_c-2)  \mu P_{n_c-2}(t) - ((n_c-1)\mu + (n_c-1)^2\nu)P_{n_c-1}(t) \\
    &\frac{d}{dt}P_{I(t)}(t) &= (n_c-1)\mu P_{(n_c-1)}(t) - n_c^2 \nu \rho\left[n_c,I(t)\right] P_{I(t)}(t) \\
    & \frac{d}{dt} I(t) &= \mu I(t) - \nu I(t)^2 \; . \label{eq:mf-limit}
\end{eqnarray}

The equations for all occupation numbers follow the same logic as before where every transition from state $y$ to $z$ (or arrows in Fig.~\ref{fig:birth-death}) corresponds to a negative term for the state $y$ and an equal positive term for the state $z$. Importantly, the transition out of the mean-field limit and back to the master equation involves a coupling function $\rho[n_c,,I(t)]$, where we need to approximate the distribution of states around the expected $I(t)$ mean-field limit. In general, we recommend using a truncated Poisson distribution. It is not the right distribution in this particular birth-death example, because of the quadratic death term, but still works relatively well. Thus $\rho[n_c,I(t)]$ is the probability that a system drawn from a Poisson distribution with mean $I(t)$ and support $[n_c,\infty]$ is found at exactly $n_c$, on the edge of that distribution and susceptible to leaving the mean-field limit.

Note that Eq.~(\ref{eq:mf-limit}) governing the position of the mean-field limit $I(t)$ is exactly the same as the standard mean-field description used in Eq.~(\ref{eq:mf}). Therefore, if we define the mean-field limit as a system with $n_c = 0$ or more active individuals, we fall back on a standard mean-field model. Conversely, as we push that mean-field limit to infinity, i.e., $n_c \rightarrow \infty$, we recover an exact master equation description. Between these limits, we approximate the true distribution of states, capturing stochastic extinctions around the absorbing state while maintaining a much smaller system of equations than in a true master equation framework.

\begin{figure}
    \centering
    \includegraphics[width=0.5\linewidth]{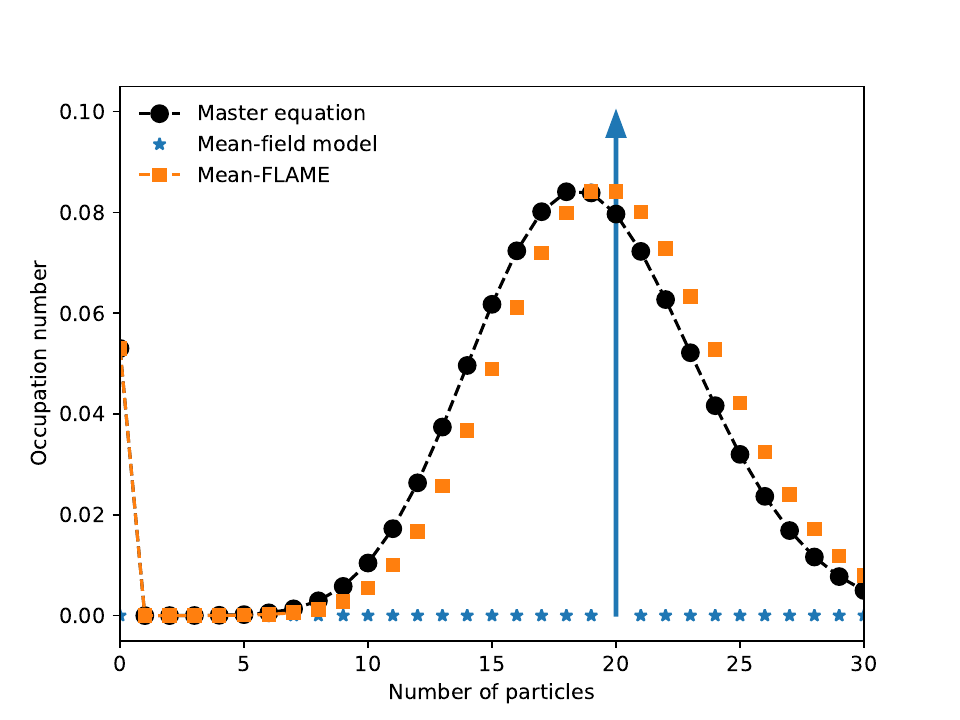}
    \caption{Birth-death process where the number of individuals $n$ in a system grows by one at rate $\mu n$ and decreases by one at rate $\nu n^2$. We set $\mu/\nu = 1/0.05 = 20$ and integrate the system up to a time $t=400$. The system is described with three approaches: A mean-field description with a single equation (occupation number of 1) leading to a steady-state at 20; an exact master equation with 100 equations, and our hybrid mean-FLAME model with 8 master equation state coupled to a mean-field limit. Markers representing the master equation states of the hybrid model are connected by a dashed line, but markers for the approximated states based on the mean-field limit are not. Note that the underlying process does not produce a Poisson distribution, such that our hybrid model is imperfect and could be improved by a better choice of $\rho[n_c,I(t)]$. Even then, the hybrid model succeeds at our main objective as it captures stochastic extinctions relatively well.}
    \label{fig:res-birth-death}
\end{figure}

We compare the predictions of the different models introduced so far in Fig.~\ref{fig:res-birth-death}.

\subsection*{Coupled subsystems with approximate master equations with mean-field limits}

Consider now that there are two weakly coupled subsystems (e.g., neighboring populations) undergoing the same birth-death process. Active individuals in one subsystem give birth to new entities that migrate to the other subsystem at a rate $\lambda$.

In a master equation framework, we would then want to track the overall state of the system using $\overrightarrow{y} = [n_1, n_2]$ capturing the number of active individuals in both subsystems. It would therefore be natural to extend this idea to our master equation with mean-field limit. However, the complexity of this approach obviously grows exponentially with the number of subsystems.

We rely on a simpler approximation to describe coupled subsystems, called an approximate master equation\cite{hebert2010propagation,marceau2010adaptive,stonge2021master}. The idea is to describe every system separately with its own master equation, and couple them using a mean-field quantity. Using this idea on master equations with mean-field limit, we can write the following system of equations for the two subsystems birth-death process, with subsystems labeled $i$ and $j$ $\in$ $[1,2]$,
\begin{align}
    \frac{d}{dt}P^{(i)}_{n}(t) &= (n+1)^2 \nu P^{(i)}_{n+1}(t) + \left[(n-1)  \mu + \theta_j\right]P^{(i)}_{n-1}(t) - (n \mu + \theta_{j} + n^2\nu)P^{(i)}_{n}(t) \\
    \frac{d}{dt}P^{(i)}_{n_c-1}(t) &= n_c^2 \nu \rho\left[n_c,I(t)\right] P^{(i)}_{I(t)}(t) + ((n_c-2)  \mu + \theta_{j}) P^{(i)}_{n_c-2}(t) - ((n_c-1)\mu + \theta_{j} + (n_c-1)^2\nu)P^{(i)}_{n_c-1}(t) \\
    \frac{d}{dt}P^{(i)}_{I(t)}(t) &= ((n_c-1)\mu + \theta_{j}) P^{(i)}_{n_c-1}(t) - n_c^2 \nu \rho\left[n_c,I(t)\right] P^{(i)}_{I(t)}(t) \\
    \frac{d}{dt} I(t) &= \mu I(t) + \theta_{j} - \nu I(t)^2 \; .
\end{align}
Where we introduced an approximate mean-field coupling representing the expected number of births in $i$ coming from $j$,
\begin{equation}
    \theta_{j} = \lambda \left[\sum_{k=0}^{n_c-1} kP_k^{(j)}(t) + I(t)P^{(j)}_{I(t)}(t)\right] \; .
    \label{eq:avg-coupling}
\end{equation}
which is the expected migration rate based on the average state of the neighbor subsystem. The average state is calculated as a sum over the number $k$ of individuals in the subsystem $j$, plus its own mean-field limit. Known correlations could be included using a clever moment closure in Eq.~(\ref{eq:avg-coupling}), such as network effects or other correlations \cite{hebert2010propagation,marceau2010adaptive}. Yet, taking a simple average is a good first-order approximation for most processes.

Importantly, this approach does not exactly capture joint extinctions where both subsystems have $0$ active individuals. For example, the state where subsystem $i$ goes extinct, $P^{(i)}_{0}$, is no longer an absorbing state due to the average coupling $\theta_j$ although it does capture the vanishing internal dynamics of system $i$. The joint absorbing state can still be reached, but is not exactly described since we ignore correlations between the subsystems such that $\theta_i$ or $\theta_j$ do not go to zero. This is the cost of using a mean-field coupling across subsystems for the resulting model to scale linearly with the number of subsystems rather than exponentially.

\begin{figure}
    \centering
    \includegraphics[width=0.5\linewidth]{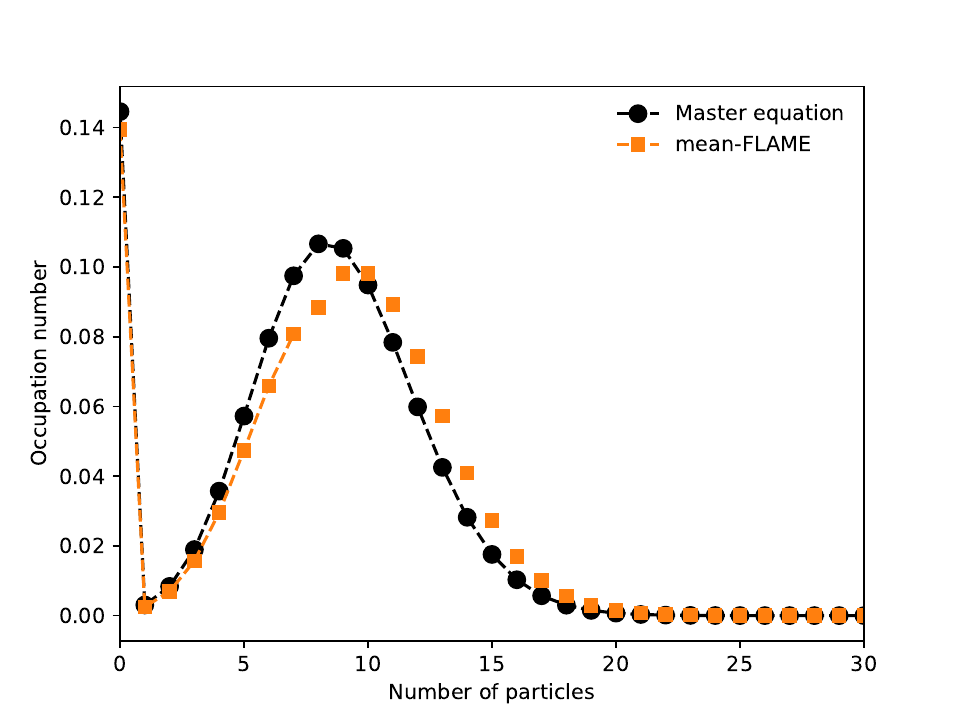}
    \caption{Two coupled birth-death processes where the number of individuals $n_i$ in system $i\in \lbrace1,2\rbrace$ grows by one at rate $\mu n_i + \lambda n_{j\neq i}$ and decreases by one at rate $\nu n_i^2$. We set $\mu = 1$, $\nu = 0.1$, and $\lambda = 0.004$ and integrate the system up to a time $t=100$. We then compare an exact master equation with 50x50=2500 equations and a mean-FLAME model consisting of two systems of 8 master equation state plus 2 states describing their mean-field limit for a total of 20 equations. Markers representing the master equation states of the hybrid model are connected by a dashed line, but markers for the approximated states based on the mean-field limit are not. Note that almost 15\% of systems would be extinct by time $t=100$ if not for the coupling mechanism, such that $\lambda = 0.004$ is far from a trivially weak coupling and still well captured by the approximate master equations.}
    \label{fig:coupled-birth-death}
\end{figure}

We compare how well mean-FLAME captures the coupled stochastic birth-death processes in Fig.~\ref{fig:coupled-birth-death}. 

\section{Two-dimensional generalization with a Lotka-Volterra example\label{sec:LV}}

Species interactions, modeled with Lotka-Volterra systems of equations, are one of the most ubiquitous examples of mean-field models in ecology. They capture a simplified version of predator-prey or competition dynamics. But in many ways, the approximations made are extreme. Consider the following predator-prey dynamics in a population of $F(t)$ prey fish and $S(t)$ sharks at time $t$. Every prey reproduces logistically at rate $\mu\left[K_F-F(t)\right]$ in an environment with carrying capacity $K_F$ (which only affects the dynamics of the prey fish, and not that of the sharks). Predators can come into contact and eat any prey at a given rate $\beta$. Predation allows the predator to have enough energy to reproduce (at rate $\beta$F), and die at a background rate, $\nu$. Assuming that contacts between predators and prey are well mixed, we can write the following mean-field system:
\begin{align}
    \frac{d}{dt}F & = \mu F\left[K_F-F(t)\right] - \beta F(t) S(t) \nonumber \\
    \frac{d}{dt}S & = \beta F(t) S(t) - \nu S(t)
    \label{eq:lv-mf}
\end{align}

These classic Lotka-Volterra equations yield some interesting insights about cycles in the population of predators and prey. However, populations in these cycles can go arbitrarily close to zero, which are absorbing states; the model does not allow extinctions because it ignores the fact that the dynamics are stochastic and that animals are discrete quantities \cite{henson2001lattice}.

In this section, we focus on a single system (no spatial aspects or coupling between subsystems) to demonstrate how to generalize the internal dynamics of mean-FLAME models to systems with more than one degree of freedom (or variable of interest).

\subsection*{Two-dimensional master equation with mean-field limit}

\begin{figure}
    \centering
    \includegraphics{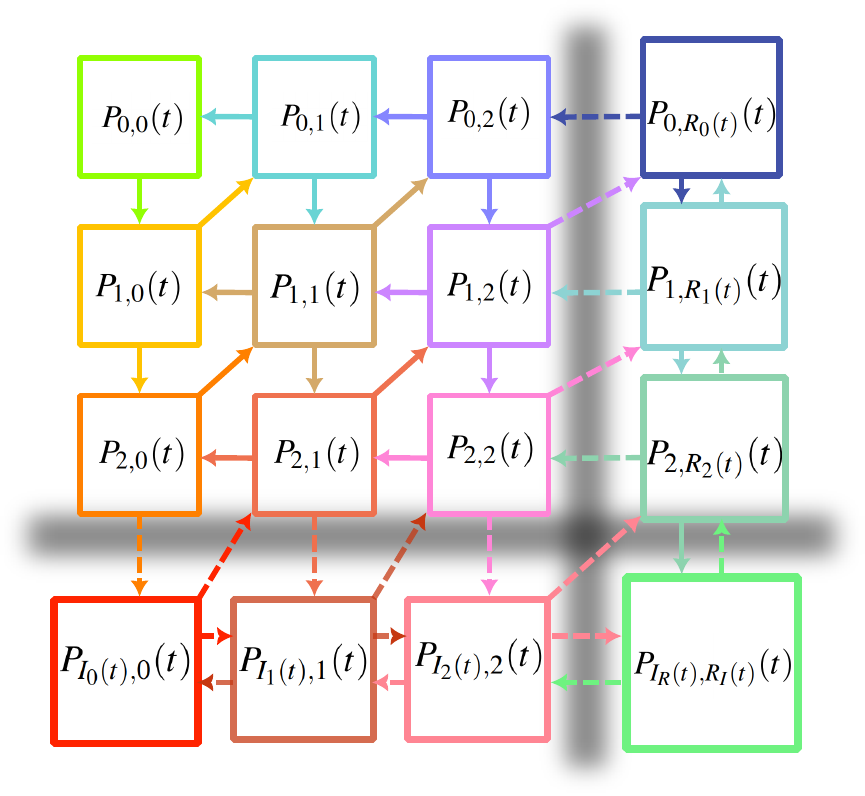}
    \caption{\textbf{Master equation with mean-field limits for Susceptible-Infectious-Recovered-Susceptible dynamics.} The exact same state and transition schema also applies to predator prey dynamics, if we simply replace the $I$ (infectious) quantities for $F$ (fish) to represent prey numbers; and the $R$ (recovered) quantities for $S$ (sharks) to represent predator numbers. Note that in both processes, the rate of the transition from $(0,0)$ to $(1,0)$ is actually zero if the system is closed such that $(0,0)$ is an absorbing state. Migration events can destabilize this absorbing state and re-introduce the infection or the prey in the system.}
    \label{fig:SIRS}
\end{figure}

We now generalize our mean-FLAME approach to multidimensional systems, which gets conceptually quite complicated. Readers can refer to the state and transition schema shown in Fig.~\ref{fig:SIRS}. Assuming the number of fish is a vertical dimension and the number of sharks is a horizontal dimension, we can explain the different transitions as follows. Vertical state transitions correspond to prey reproduction while transitions to the left correspond to predator death. The act of predation, in the general master equation states, corresponds to the diagonal transitions that simultaneously decreases the number of prey by one and increases the number of predator by one. However, in the mean-field limit, this transition is diagonal only if the number of prey is such that the systems lies at the edge of the mean-field approximation and predation takes the prey system out of its limit. Otherwise, it is a horizontal move to the right that increases the number of predator while also leaving the prey system in the mean-field limit of prey numbers.

We now have to introduce many mean-field limits for both prey and predator numbers given all possible combinations of population sizes for prey and predator populations. There is a mean-field quantity for the number of prey given zero predators in the system, or one predator, or two predators, and so on; likewise for mean-field limits for the number of predators. Finally, there is a double mean-field limit where counts of both predator and prey are approximated by mean-field quantities.

In two dimensions, we therefore have 9 types of occupation numbers: 1) The general master equation, 2-3) states that are next to one mean-field limit but not the other, 4) a single state that is next to both mean-field limits, 5-6) mean-field limit in one dimension but far from the limit in the other, 7-8) mean-field limit in one dimension and next to the other limit, and 9) the double mean-field regime. This is without counting the mean-field quantities that also need to be followed.

First, the general master equation of a Lotka-Volterra predator-prey system can be written like so
\begin{align}
\frac{d}{dt}P_{n_F,n_S}(t) = -&\mu n_F(K_F-n_F)P_{n_F,n_S}(t) -\nu n_S P_{n_F,n_S}(t) - \beta n_F n_S P_{n_F,n_S}(t)
     + \mu (n_F-1)(K-n_F+1)P_{n_F-1,n_S}(t) \nonumber \\
     &+ \nu (n_S+1) P_{n_F,n_S+1}(t) + \beta (n_F+1)(n_S-1) P_{n_F+1,n_S-1}(t) \; .
\end{align}
For large numbers of prey and a small number of predators, the system is eventually coupled to a mean-field limit. If the limits start at $n_F = n_{cF}$ and $n_S = n_{cS}$, we write the following close to the limit in $n_F$ but with $n_S<n_{cS}-1$:
\begin{align}
\frac{d}{dt}P_{n_{cF}-1,n_S}(t) = -&\mu (n_{cF}-1)(K_F-n_{cF}+1)P_{n_{cF}-1,n_S}(t) -\nu n_S P_{n_{cF}-1,n_S}(t) - \beta (n_{cF}-1) n_S P_{n_{cF}-1,n_S}(t) \nonumber \\
     &+ \mu (n_{cF}-2)(K_F-n_{cF}+2)P_{n_{cF}-2,n_S}(t)
     + \nu (n_S+1) P_{n_{cF}-1,n_S+1}(t) \nonumber \\
     &+ \beta n_{cF} (n_S-1) \rho\left[n_{cF},F_{n_S-1}(t)\right]P_{n_{cF},n_S-1}(t) \; .
\end{align}
A similar equation governs the occupation number of the corresponding mean-field limit in the number of prey:
\begin{align}
\frac{d}{dt}P_{n_{cF},n_S}(t) =  -&\nu n_S P_{n_{cF},n_S}(t) - \beta F_{n_S}(t) n_S P_{n_{cF},n_S}(t)
     + \mu (n_{cF}-1)(K_F-n_{cF}+1)P_{n_{cF}-1,n_S}(t) \nonumber \\
     &+ \nu (n_S+1) P_{n_{cF}-1,n_S+1}(t) + \beta  F_{n_S-1}(t) (n_S-1) \lbrace1-\rho\left[n_{cF},F_{n_S-1}(t)\right]\rbrace P_{n_{cF},n_S-1}(t) \; .
\end{align}
Notice that some subtle quantities appear in these equations, such as the mean-field quantity for prey, $F_{n_S-1}(t)$, which corresponds to the mean-field expectation for the number of prey fish in a system with $n_S - 1$ sharks at time $t$. That quantity in the predation term as the mean-field quantity itself mediates the rate at which predation occurs in the mean-field limits. The important distinction in structure with previous equations is that predation only adds to this equation number if the system in which predation occurs does not leave the mean-field limit (diagonal versus lateral move); hence the $1-\rho$ term. These last equations are then a closed system once we write out the corresponding mean-field quantities,
\begin{equation}
    \frac{d}{dt}F_{n_S}(t) = \mu F_{n_S}(t)\left[K_F - F_{n_S}(t)\right] - \beta n_S F_{n_S}(t) \; .
\end{equation}

The same approach can be used to describe the mean-field limits of the number of predators. The full system of equations for this case study is presented in our supplemental appendix, along with all of our other case studies.

\begin{figure}
    \centering
    \includegraphics[width=0.5\linewidth]{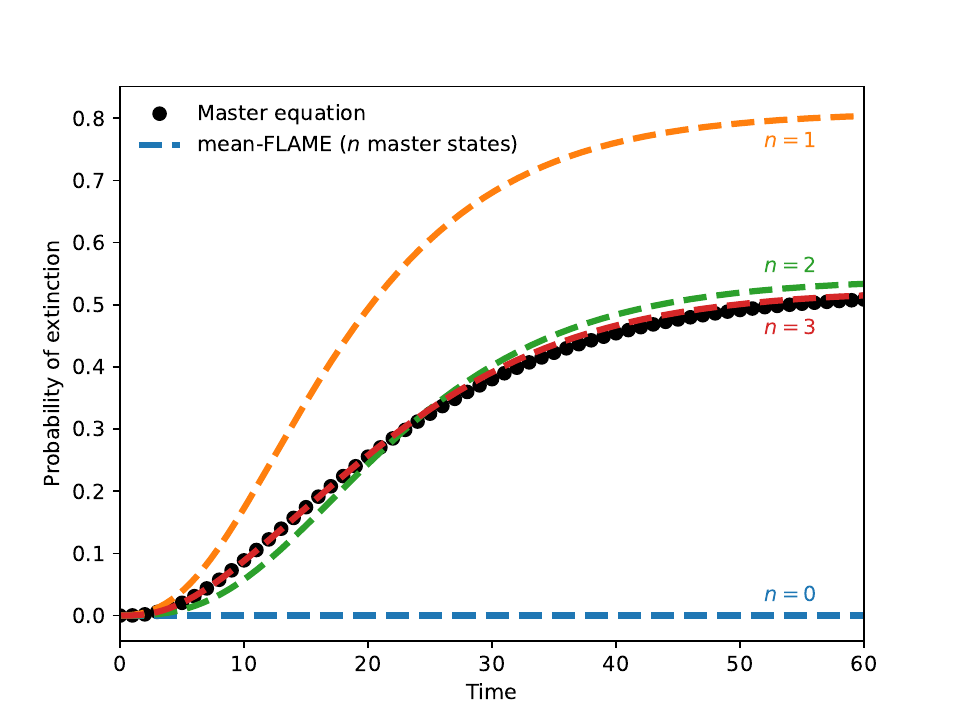}
    \caption{Extinction probability in a Lotka-Volterra model where the system starts with 1 prey and 1 predator, then follows the dynamics described in the main text with $\mu = 0.005$, $\nu = \beta = 0.1$ and a carrying capacity of $K_F=20$. The full master equation systems requires $(K_F+1)^2 = 441$ equations. We then compare predictions for the extinction probability over time using master equation with mean-field limit and a variable number of master equation states $n$ per species (i.e., we explicitly track states with less than $n$ individuals of a certain type). We fall back on the classic mean-field Lotka-Volterra model by using $n=0$, and find that it of course cannot predict extinction. However, we almost perfectly recover stochastic extinction starting at $n=3$. At this point, the mean-FLAME system consists of $n\times n = 9$ master equation states and $2n + 1 = 7$ mean-field limits. The schematic of that model can be found in Fig.~\ref{fig:SIRS}.}
    \label{fig:lotka-volterra}
\end{figure}

\subsection*{Extinction probability}

To illustrate the usefulness of the approach, we experiment with the ability of the master equations with mean-field limits to correctly recover the probability of extinction in a Lotka-Volterra system. We define extinction as a system with no prey or predator left by time $t$, and the probability of that happenning is captured by $P_{0,0}(t)$ in the above equations. The results of this case study are shown in Fig.~\ref{fig:lotka-volterra}. We use systems with $n$ explicit master equation states per species. With $n=0$, we have a fully mean-field description which is deterministic and thus predict no extinctions. Interestingly, a system with $n=1$ actually overpredicts extinctions, most likely because the joint-Poisson approximation for the double mean-field regime underestimates correlations and overestimates the likelihood of having low numbers of both predator and prey. That being said, the system gives near perfect predictions with $n$ as low as three. In this particular case, compared to a full master equation framework, that is 20 times fewer equations in the mathematical model without much loss of precision.

\section{Metapopulation epidemic models\label{sec:epi}}

Having introduced all of our modeling techniques, we now turn our focus to the two spatially explicit complex models described in the introduction. For the first example, we will show how our proposed technique can be used to track the spread of an emergent pathogen through a heterogeneously distributed metapopulation where most local communities are very sparsely populated.

\subsection*{Susceptible-Infectious-Recovered-Susceptible dynamics}

In this example we want to model an epidemic process where individuals can take one of three states: susceptible ($s$), infected ($i$), or recovered ($r$). Individuals can cycle through all states and back based on a general set of mechanisms referred to as the SIRS process\cite{anderson1991infectious}. Susceptible individuals become infected by contact with infected individuals at a infection rate $\beta$. Infected individuals recover and gain immunity at a recovery rate $\alpha$. And recovered individuals lose their immunity and become susceptible again according to a waning rate $\gamma$. In general, the epidemic can be summarized through its reproduction number $R_0 = \beta N/\alpha$ which represents the expected number of infections caused by a first infected case. An epidemic would die out with $R_0<1$, but grow exponentially when $R_0>1$. We can then also define a herd immunity threshold generally estimated as $h = 1-1/R_0 = 1-\alpha/\beta N$ such that $h$ is the fraction of the population that needs to be immune for the reproduction number $R_0 = \beta N(1-h)/\alpha$ of the infectious disease to drop below one in a population with $N(1-h)$ susceptible individuals. If there are more susceptible individuals, the section of the population of immune individuals is below the herd immunity threshold and can cause a supercritical epidemic (defined as $R_0>1$) until the the herd immunity threshold is reached.

Current epidemic models of this process that are amenable to simpler mathematical techniques work best in large, densely populated (e.g., large population centers) population centers with well-mixed contact patterns between individuals. However, in sparsely populated areas (e.g., rural areas) these methods are less effective since the influence of small population numbers and stochastic fluctuations can have greater impact and need to be considered more explicitly.

In fact, stochastic effects can have surprising impacts in small populations which are never considered in models used for large, high density populations. For example, consider a stochastic epidemic process that starts in a small local population of 50 individuals. Even with $R_0>1$, transmission might sputter out quickly \cite{hebert2020beyond} and die out as a small cluster of 5 cases, for example. The settlement is then left with 45 susceptible people, a 10\% reduction in global susceptibility. Small populations can therefore easily reach herd immunity without ever seeing supercritical outbreaks since even small outbreaks can build immunity in a large fraction of the population. This can in turn affect how easily or quickly an epidemic might be able to spread across a sparsely populated area.

In a population of size $N$, we need to know the size of the number of individuals in at least two out of three epidemiological states to know the complete state of the system. For example, if we know the numbers $i$ and $r$ of infectious and recovered individuals, we know that we have $s=N-i-r$ susceptible individuals given that we have a closed population. We can therefore use a two-dimensional mean-FLAME approach to describe the system. 

The approximate master equation for a single SIRS process without mean-field limits is given by
\begin{align} \label{eq:master_eq_mean_field_sir}
    \frac{d}{dt}P_{i,r}(t) = & -(\beta i+\theta)(N-i-r) P_{i,r}(t) - \alpha i P_{i,r}(t)  - \gamma r P_{i,r}(t) \nonumber \\  
    & + (\beta(i-1)+\theta)(N-i-r+1)P_{i-1,r}(t) + \alpha(i+1)P_{i+1,r-1}(t) + \gamma(r+1)P_{i,r+1}(t)
\end{align}
where $\theta$ is again the mean-field coupling estimate for one local population to other populations of the metapopulation. We then introduce two-dimensional mean-field limits for both the $i$ and $r$ dimensions to simplify the description of large epidemics in large populations. We can denote these $I_r(t)$ (or $R_i(t)$) for mean-field numbers of infectious (or recovered) individuals in populations with exactly $r$ recovered (or $i$ infectious) individuals; and $P_{I_r(t),r}(t)$ (or $P_{i,R_i(t)}(t)$) represent the occupation probabilities of these states. For populations in the double mean-field limits, we track $I_{R(t)}(t)$ and $R_{I(t)}(t)$. The schema of the model is presented in Fig.~\ref{fig:SIRS} and the resulting system of equations is again written in supplemental appendix. 

To account for spatially distributed metapopulations of potentially infected individuals, we embed $n$ by $n$ such systems in space where every local population is given a spatial position $(x,y)$, a set of states as in Fig.~\ref{fig:SIRS}, and a connectivity kernel to other local populations $j \in \mathcal{N}_{(x,y)}$ which represents neighboring populations. The set of all neighborhoods $\mathcal{N}_{(x,y)}$ could be any arbitrary metapopulation network. Over these coupled metapopulations, we can calculate the expected coupling $\theta^{(x,y)}$ specific to the $(x,y)$ local population as
\begin{equation}
    \theta^{(x,y)}  = \lambda \beta \sum_{j\in\mathcal{N}_{(x,y)}} \bigg\lbrace \sum_{i,r} iP^{(j)}_{i,r}(t) + \sum_r I_r(t)P^{(j)}_{I_r(t),r}(t) + I_{R(t)}(t)P^{(j)}_{I_R(t),R_I(t)} \bigg\rbrace \; .
\end{equation}
This mean-field quantity is again a basic coupling term with a $\lambda$ factor (akin to a migration rate). This factor multiplies a sum over all neighbors $j$ of the $(x,y)$ local population where we calculate their expected number of infected individuals by summing over the three types of states: general master equation (first sum in braces), infectious mean-field with exact recovered state (second sum in braces) and double mean-field state (final term in braces). Note that this mean-field coupling does not depend on the state of the $(x,y)$ local populations, which makes calculations easier but means that we ignore all correlations induced by the dynamics. That being said, a future version of the model could include known correlations in this coupling, since if we know that the population in $(x,y)$ is in a state with low or high infectious counts, then neighboring populations are likely to be in a similar state.

\subsection*{Spread of epidemics in heterogeneous spatially-distributed populations}

\begin{figure}
    \centering
    \includegraphics[width=0.5\linewidth]{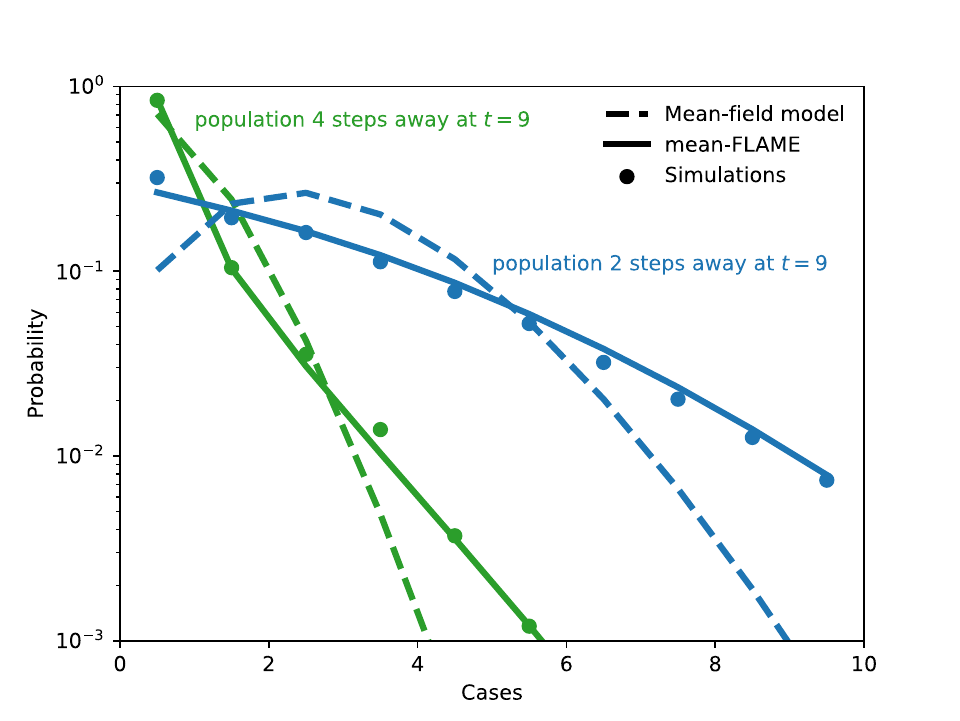}
    \caption{Distribution of cases per population in a SIRS metapopulation model. We use a finite system of populations at specific $(x,y)$ positions with $x$ and $y \in [0,50]$ and with the local populations equal to $N_{(x,y)}=\ceil{10+100(\cos{x/5})^2(\sin{\sqrt{y/2}})^2}$. Populations are fully connected internally, and to their direct neighbors at populations $(x+1,y)$, $(x-1,y)$, $(x,y+1)$ and $(x,y-1)$ if they exist. We then run a SIRS process with $\beta = 0.025$, $\alpha = 1$, $\gamma =0.01$, and $\lambda = 0.1$ with initial conditions of one infectious cases in the 25 populations with $x<5$ and $y<5$. We evaluate the distribution of cases by time $t=9$ for two specific populations a certain number of steps away from the nearest initial case. The mean-FLAME model is integrated using 10 exact states for both infectious and recovered individuals, across all 50 by 50 locations this results in a total of 357,500 equations. The schematic of the system can be found in Fig.~\ref{fig:SIRS}. The mean-field model uses up to 7,500 equations. A complete approximate master equation model would take up to $\textrm{max}(N_{(x,y)})^2$ per location, so up to about 30 million equations. A true master equation approach is intractable. We compare to the distribution obtained from 10,000 simulations of the same process with an exact algorithm \cite{st2019efficient}.}
    \label{fig:sirs-time}
\end{figure}

We evaluated how well a mean-FLAME model captured the stochastic diffusion of an SIRS epidemic process through a metapopulation system by comparing it to a classic mean-field description. We quantified the temporal patterns of stochatistic diffusion by looking at the time to introduction of the epidemic in specific local populations. The results of this case study are shown in Fig.~\ref{fig:sirs-time}.

As expected, a classic mean-field description overestimates the homogeneity of potential disease spread. We obtain this mean-field description by tracking the expected number of infected and recovered individuals for each local population in the metapopulation system, and then by assuming a Poisson distribution around that average. Doing so creates two big issues. First, the mean-field approach overestimates how likely it is for a given population to have had a first infected individual. In other words, even if the expected number of infected in a given local population is less than one, internal dynamics drive the expected number of infections to increase when it should not. Second, the mean-field approach assumes that the number of internal contacts between susceptible and infectious individuals is given by the product of the expected numbers of susceptibles and infectious individuals, ignoring potential correlations between these two numbers. This slows down the spreading dynamics for systems in the tail of the distribution of actives cases. If the number of active cases is higher than average, the number of recovered cases might be lower than average since some cases are taking longer to recover, leading to a higher total number of contacts between susceptible and infected. However, mean-FLAME does not suffer from this same issue. This allows the mean-FLAME model to produce more heterogeneous distributions of potential cases per population than classic mean-field models (see Fig.~\ref{fig:sirs-time}).

Of note, a full approximate master equation would have 12,321 equations (population size squared) per local population to capture the full range of potential number of infectious and recovered cases. A real master equation framework would essentially be impossible, with a total number of equations scaling exponentially with the number of populations. Yet, the mean-FLAME model used in Fig.~\ref{fig:sirs-time} runs more efficiently by using only 142 equations per population by approximating high numbers with a mean-field limit.

\begin{figure}
    \centering
    \includegraphics{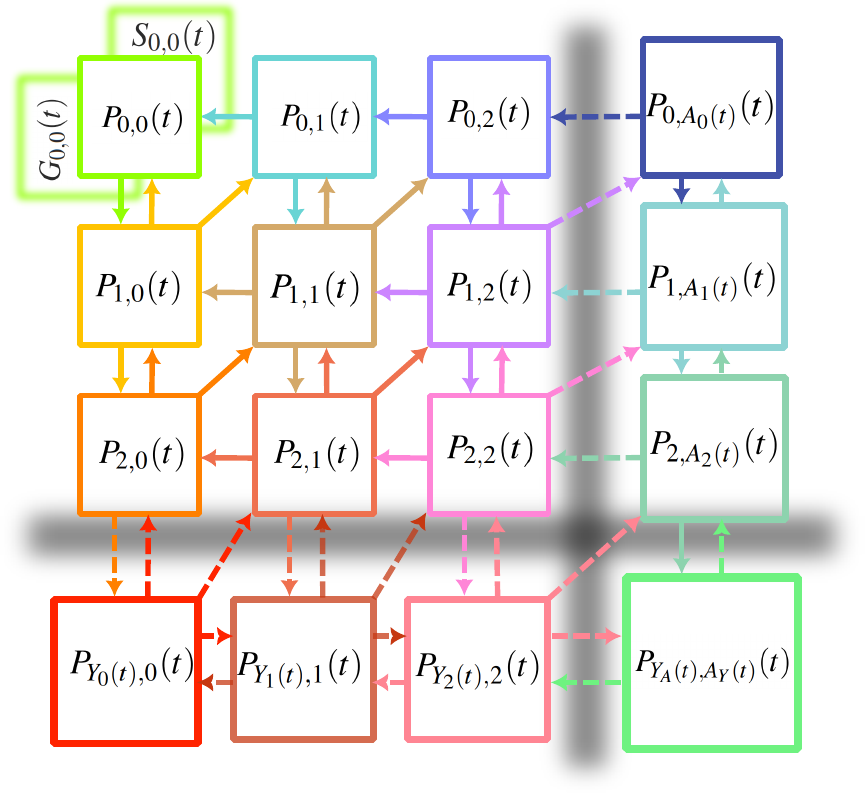}
    \caption{\textbf{Master equation with mean-field limits for tree species life-cycle.} We track the population of trees over four stages of their life-cycle: S seeds (state $S$), seedlings (state $G$ for greens), saplings (state $Y$ for young trees) and adult trees (state $A$). The last two stages are important stochastic bottlenecks in the growth of a population and therefore captured by master equations with mean-field limits. The horizontal dimension of the schema represents the number of adult trees, and the vertical dimension the number of young trees. The seed and seedling dimensions are captured as a fully mean-field description (no explicit master equation state) and shown for the absorbing state only for simplicity, but note that these dimensions exist for all states including mean-field limits. Across these states, we capture arrival of new saplings and death of saplings (vertical arrows), growth of saplings into adult trees (diagonal arrows) and death of adult trees (horizontal arrows). }
    \label{fig:life-cycle}
\end{figure}

\section{Species dispersal model\label{sec:dispersal}}

For our final example, we approximated the range shift of a tree in response to a slowly shifting landscape of suitable environments driven by climate change. Specifically, we model a tree species that moves by seed dispersal over a discrete space $(x,y)$ where every grid cell in that space has local conditions that specifies the cell's local carrying capacity $K(x,y,t)$ for the tree species at time $t$.

\subsection*{Handling of complex life cycles}

We increased the realism of our model by considering a four-stage life cycle for the tree species: (i) Seeds ($s$), (ii) seedlings or small greens ($g$), (iii) saplings or young trees ($y$), and (iv) adult trees ($a$). Each stage dies at a rate, $\nu_x$, and transitions into the next stage of the cycle at rate $\mu_x$, where all rates are stage dependent. For instance, adult trees produce seeds at a fast rate of $\mu_a$ while young trees transition into adults at a slower rate equal to $\mu_y$ times the square root of the number of young trees times the difference between the carrying capacity and the current population. This transition from young to adult tree stages capture the non-linear growth expected when increasing biomass leads to an increase in competition for resources among individuals \cite{hatton2024diversity}.  The inclusion of non-linear terms in a model increases the potential error of using deterministic mean-field models. Since the population of trees at any stage of the life cycle is unconstrained, this model has four degrees of freedom for each spatial cell.

However, instead of systematically moving to master equation models of higher dimensionality, we can also use purely mean-field descriptions with no explicit master equation states for degrees of freedom that are systematically far from absorbing states. For example, when modeling a tree species, the number of seeds S(t) might not need to be explicitly tracked as it tends to be high and is therefore not as sensitive to fluctuations near zero. This allows us to focus on stages of the life cycles which are population \textit{bottlenecks}. In this example, the numbers of young trees $Y(t)$ and adult trees $A(t)$ are more important to track explicitly as they are fewer in numbers, have more influence, and drive the numbers of seeds and seedlings at lower stages of the life cycles. We therefore track $Y(t)$ and $A(t)$ with master equations and mean-field limits, and these quantities will influence the purely mean-field descriptions for the number of seeds $S_{Y(t),A(t)}(t)$ and seedlings or small greens $G_{Y(t),A(t)}(t)$. We can think of this generalization as adding a third and fourth dimension to our existing model, but one where there are no explicitly tracked master equation states, only mean-field equations. The resulting state transition scheme is illustrated in Fig.~\ref{fig:life-cycle} for a tree life-cycle and the full system of equations can be found in our supplemental appendix.

\subsection*{Speed of diffusion across changing climate}

\begin{figure}
    \centering
    \includegraphics[width=0.9\linewidth]{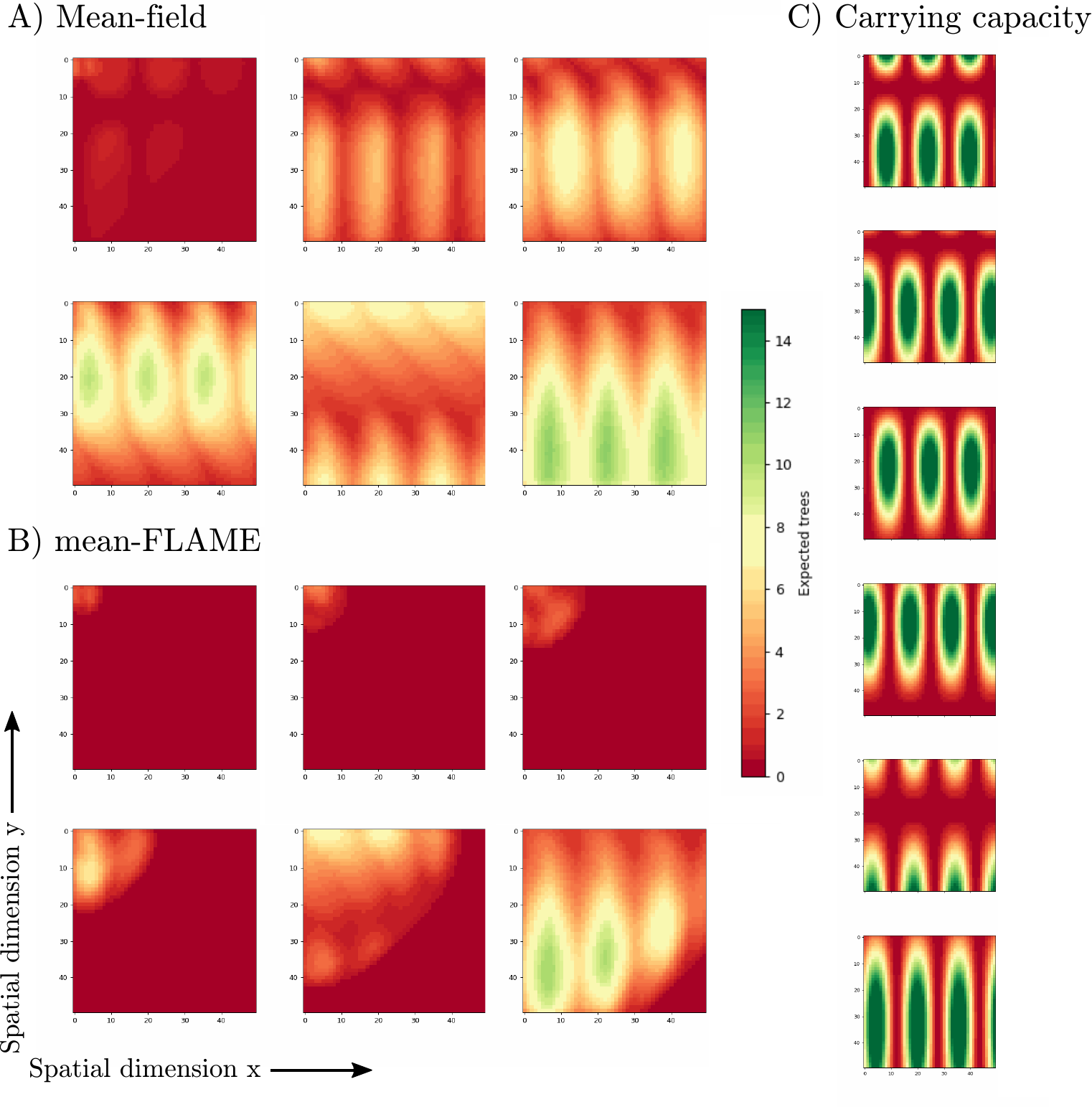}
    \caption{We run the diffusion of a tree species with a four-stage life cycle as described in the main text using both mean-field and mean-FLAME models (with 5 exact states for young and adult trees alike). The behavior shown is that of a typical run regardless of parameters, which are chosen around the following values: $\mu_a = 1000.0$, $\mu_s = 0.5$, $\nu_s = 20.0$, $\mu_g = 0.2$, $\nu_g = 1.0$, $\mu_y = 0.1$, $\nu_y = 0.5$, and $\nu_a = 0.05$. We start the dynamics with no trees but 10 seeds and 20 seedlings in all sites with $x<5$ and $y<5$. The entire landscape is a grid of $(x,y)$ with $x$ and $y$ from 0 to 50, definition locations with local changing climate modeled by a carrying capacity $K(x,y,t)=\vert{20(\sin{\sqrt{(x+t)/2}})^2(\cos{(y+t)/5})^2}$. Seeds are distributed locally and to the von Neumann neighborhood of a given site. We then track the dynamics from $t=0$ to 100 and output the expected number of adult trees based on both approaches at time $t=7.5$, 15, 22.5, 30, 50, 90 (respectively, going from the top left to top right, then bottom left to bottom right panels). The deterministic mean-field approach dramatically overestimates the initial speed at which the tree species can explore the landscape. Note that a standard approximate master equation approach is intractable in this case (too many equations) and mean-field limits are therefore required. Finally, the time-varying local carrying capacity is shown in panel C) for comparison.}
    \label{fig:gif-movie}
\end{figure}

In Fig.~\ref{fig:gif-movie}, we show an example of how a deterministic mean-field approach can overestimate the speed of diffusion of a species over a landscape. The key factor is that the mean-field approach does not capture non-linear terms correctly. For example, sites with very low carrying capacity might spend most of their time without adult trees yet will consistently produce seeds in the mean-field version. Likewise, we used a model where young trees are competing for resources (e.g, light) such that the growth potential of the population scales with the square-root of the number of young trees rather than linearly. The mean-FLAME approach captures the fact that young trees are discrete quantities and systems will have 0 young trees, or 1, or 2, etc. In a mean-field model, the growth rate of the first young trees is artificially boosted when their expected quantity can be any continuous number above zero.

Finally, note that the results of Fig.~\ref{fig:gif-movie} used four mean-field equations per grid cell in the mean-field model, and a total of 141 equations per grid cell in the mean-FLAME implementation. The latter is composed of 5x5=25 exact master equation states describing the number of young and adult trees, followed by 10 mean-field limits (two equations each) and 1 double-mean-field regime (two equations), multiplied by 3 to add mean-field limits of seeds and greens. A similar but smaller mean-FLAME implementation is represented in Fig.~\ref{fig:life-cycle}.

\section{Conclusion}

Standard mathematical models that rely on deterministic or mean-field approximations struggle to capture scenarios where we need to account for the stochastic and discrete nature of spreading or diffusion processes. We developed a master equation description close to absorbing states coupled with mean-field limits far from the absorbing states. This hybrid mean-FLAME modeling approach allows us to succinctly capture local extinctions mechanisms as well as stochastic bottlenecks in life cycles or spreading processes. In a nutshell, when tracking only $\epsilon$ states around an absorbing state in a $d$ dimensional system of $m$ populations of size $N$, mean-FLAME appears to perform as well as a full approximate master equation while requiring $m\epsilon^d$ equations instead of $mN^d$ equations. That represents a gain of $(N/\epsilon)^d$ in computational complexity. Conversely, a full master equation approach is often impossible as it would require a $N^{md}$ equations. This speed-up can be incredibly useful in high-dimensional systems.

Implementation of our methods is not without challenges. First, as made clear in our case studies, one must carefully walk through multiple types of states. There are 3 types of states when the system has a single dimension---the general master equation, the state coupled to the mean-field limit, and the mean-field limit---but $3^d$ types of states in $d$-dimensional systems. Second, numerical integrators might struggle with a single system of equations where some values like the occupation number have small values and derivatives, while others like the position of the mean-field limits can have gigantic values. Interestingly, the mean-field limits behave independently of their occupation numbers such that one can integrate the mean-field equations first and then use their solutions as an input to integrate the system of occupation numbers. As a starting point, we offer different numerical approaches in both C++ and python as part of our software release. Some pedagogical examples are available at \url{https://bit.ly/meanFLAME} and we have a complete repository of our different implementations at \url{https://github.com/nwlandry/mean-FLAME}.

Future work could improve on the presented hybrid approach by creating stages of mean-field limits, essentially binning non-explicit states in different ways. Going back to a birth-death process, the number of active individuals could diverge if the death rate was not quadratic and we could then imagine creating a mean-field buffer for systems with more than 10 individuals but less than 100 and a mean-field limit beyond that. Following a master equation description with a series of mean-field compartments with carefully calibrated coupling might help capture the heterogeneity of growth processes that do not converge to a homogeneous distribution around a steady-state (e.g. rich-get-richer processes).

Recall that the mean-FLAME approach introduced here can fall back on our standard mean-field models in one limit (mean-field limit going to the absorbing state) and on standard master equations in the other (mean-field limit going to infinity). And despite ongoing implementation challenges, the resulting modeling tool can streamline the process of quickly exploring scenarios or parameter space using a simple mean-field model, then adding precision around important stochastic events as desired by adding explicit states as needed and tuning the position of the mean-field limit. We therefore hope that this modeling approach will be used to explore possible interventions to better control complex systems around their absorbing states, especially in rural or marginal areas where system sizes can be much smaller.

Finally, we have presented use cases of the mean-FLAME approach that focused on extinction dynamics, the spread of infectious diseases, and dispersal of individuals into previously unoccupied areas. However, it is worth considering other contexts that would benefit from our near absorbing state approximation of stochastic processes. Specifically, our approach might allow more explicit population viability analysis particularly for small, isolated local populations found in increasingly fragmented landscapes \cite{reed2002emerging}. Moreover, in the coming decades we expect a reshuffling of most ecological systems. Modeling techniques like we have presented here that simplify the number of states to be tracked while maintaining the multidimensional complexity of the system may improve our ability to consider the consequences of new combinations of organisms, potential priority effects that may influence the resiliency of the new ecosystem, or even the potential for new pandemics \cite{carlson2022climate}. More generally, we hope that the approach we have proposed helps to highlight critical gaps in how we are approaching the development of ecological models when we assume deterministic dynamics or that the expected state of a system represents its full distribution of possible states. Life at the margins of systems may in fact be more critical when modeling complex systems and processes.

\paragraph{\textbf{Data availability statement}} Code used to generate these results is available on \href{https://github.com/nwlandry/mean-FLAME}{GitHub} and at Ref.~\cite{hebert-dufresne_mean-FLAME_2024}

\paragraph{\textbf{Acknowledgements}} This research was supported by the US National Science Foundation (award EPS-2019470) L.H-D. also acknowledges financial support from the National Institute of General Medical Sciences of the NIH under the 2P20GM125498-06 Centers of Biomedical Research Excellence and under the 3P20GM103449-23S1 IDeA Networks of Biomedical Research Excellence.

\end{document}